# Generalized I-MMSE for $K$-User Gaussian Channels

Samah A. M. Ghanem, *Senior Member, IEEE*

*Abstract*—In this paper, we generalize the fundamental relation between the mutual information and the minimum mean squared error (MMSE) by Guo, Shamai, and Verdu [1] to K-User Gaussian channels. We prove that the derivative of the multiuser mutual information with respect to the signal to noise ratio (SNR) is equal to the total MMSE plus a covariance term with respect to the cross correlation of the multiuser input estimates, the channels and the precoding matrices. We shed light that such relation is a generalized I-MMSE with one step lookahead and lookback, applied to the Successive Interference Cancellation (SIC) in the decoding process.

## I. INTRODUCTION

Duncan, in [2] showed that for the continuous-time additive white Gaussian noise (AWGN) channel, the filtering minimum mean squared error is twice the input output mutual information for any underlying signal distribution. This has illuminated intimate connections between information theory and estimation theory which has been emphasized by Guo, Shamai, and Verdu in a seminal paper [1]. More specifically, Guo et al. have shown that in the classical problem of information transmission through the conventional AWGN channel, the derivative of the mutual information with respect to the SNR is equal to the smoothing minimum mean squared error; a relationship that holds for scalar, vector, discrete-time and continuous-time channels regardless of the input statistics. The relevance of these recent connections comes from the fact that mutual information and MMSE are two canonical operational measures in information theory and estimation theory. Later Palomar and Verdu extended this relation to linear vector Gaussian channels [3], [4]. The mutual information was also represented as an integral of a certain measure of the estimation error in Poisson channels [5], [6]. Most recently, Ghanem in [7], [8], derived the gradient of the joint, conditional and non-conditional mutual information with respect to arbitrary parameters of the multiple access Gaussian channels, a relation that extends the case of mutually interfering inputs in linear vector Gaussian channels to the case of multiple non-mutually interfering inputs and with mutual interference, a starting point to the results in this work. The results of this work extends and deepens the two-user case simple framework presented in [9] to a $K$-user framework with a comprehensive interpretation to the interference and the successive interference cancellation in the decoding process. The implications of a framework involving key quantities in information theory and estimation theory are countless both from the theoretical [10], [11] and the more practical perspective, see e.g. [12], [13], [14], [15]. For instance, most recently, Ghanem in [16] proved the existence of such connections on network-coded flows over noisy networks, opening a new horizon of engineering design of precoding and decoding solutions adapted to network topology awareness. Furthermore, connections between information measures and estimation measures allow for finding explicit closed form expressions of the mutual information for binary inputs, particularly ones for BPSK and QPSK over the single input single output (SISO) channel, [1], [17], [18].

The author in [19] provides a new look - that relies on the user decoding order -towards finding such closed forms for the multiuser case. Therefore, it is of particular importance to address connections between information theory and estimation theory for the multiuser ($K$-users) case. First, it allows for interpreting the interference with joint estimation or along the SIC decoding process of users. Second, it goes aligned with and beyond frameworks with mixtures of processes within its application to the communications framework. Third, it allows for finding schemes that are interference aware.

In this paper, we first revisit the connections between the mutual information and the MMSE for the $K$-users Gaussian channels, see also [8], [15]. Therefore, the fundamental relation between the derivative of the mutual information and the MMSE, known as I-MMSE identity, and defined for point to point channels with any noise or input distributions in [1] is not anymore suitable for the multiuser case. Therefore, we generalize the I-MMSE relation to the multiuser case, interested reader can refer to the extended version in [19].

Throughout the paper, the following notation is employed, boldface uppercase letters denote matrices, lowercase letters denote scalars. The superscript, $(.)^{-1}$, $(.)^T$, $(.)^*$, and $(.)^\dagger$ denote the inverse, transpose, conjugate, and conjugate transpose operations. The $\mathbb{E}[.]$ denotes the expectation operator. The $||.||$, $Tr\{.\}$, and $det(.)$ denote the Euclidean norm, the trace of a matrix, and the determinant of a matrix, respectively.

The rest of the paper is organized as follows, Section II introduces the system model. Section III introduces the new fundamental relation between the multiuser mutual information and the MMSE. Section IV provides the conditional and non-conditional components of the $K$-user I-MMSE identity, where a precise SIC decoding process defines per-user components in information measures and estimation measures. Section V introduces the application of the unveiled relation for the special case of Gaussian distributed inputs. Section VI provides future directions in line with the paper result. Section VII concludes the paper.

## II. SYSTEM MODEL

Consider the deterministic complex-valued multiple access Gaussian channel,

$$\mathbf{y} = \sum_{k=1}^{K} \sqrt{snr} \, \mathbf{H}_k \mathbf{P}_k \mathbf{x}_k + \mathbf{n}, \quad (1)$$

where the $n_r \times 1$ dimensional vector $\mathbf{y}$ and the $n_t \times 1$ dimensional vectors $\mathbf{x}_1, .., \mathbf{x}_K$ represent, respectively, the received vector and the independent arbitrary or continuous distributed zero-mean unit-variance transmitted information vectors from each user $k = 1, .., K$ input to the multiuser channel. The distributions of the inputs are not fixed, not necessarily Gaussian nor identical. The $n_r \times n_t$ complex-valued matrices $\mathbf{H}_1, .., \mathbf{H}_K$ correspond to the deterministic channel gains for the $K$-users input channels (known to both encoder and decoder) and $\mathbf{n} \sim \mathcal{CN}(\mathbf{0}, \mathbf{I})$ is the $n_r \times 1$ dimensional complex Gaussian noise with independent zero-mean unit-variance components. The $n_t \times n_t$ $\mathbf{P}_1, .., \mathbf{P}_K$ are precoding matrices that do not increase the transmitted power

Notice that $\mathbf{H}_k \mathbf{P}_k$ matrix can correspond to any measurement matrix $\mathbf{M_k}$ of a system of matched mixtures of arbitrary processes contaminated by AWGN.

### A. The I-MMSE Identities

In this subsection, we recap on the I-MMSE relation available. The two main results are related to scalar channels [1] and point to point vector Gaussian channels [3]. Therefore, we introduce two special cases of the system model in (4) to show each case and to understand how this paper generalizes both results.

One addressed special case of the system model is the simplest scalar point to point Gaussian channel,

$$\mathbf{y} = \sqrt{snr} \mathbf{x}_k + \mathbf{n}, \quad (2)$$

Let the signal to noise ratio (SNR) of the channel be denoted by $snr$. Both the input-output mutual information and the MMSE are monotone functions of the SNR, denoted by $I(snr)$ it has been shown by Guo, Shamai, Verdu in [1] that for point-to-point Gaussian channels, the derivative of the mutual information with respect to the SNR is equal the MMSE[1], regardless of the input statistics,

$$\frac{dI(snr)}{dsnr} = mmse(snr) \quad (3)$$

The second addressed system model is for point to point vector Gaussian channel,

$$\mathbf{y} = \sqrt{snr} \, \mathbf{H}_k \mathbf{P}_k \mathbf{x}_k + \mathbf{n}, \quad (4)$$

For such vector channel, the gradients where derived with respect to different arbitrary parameters. However, the theorem of Guo, Shamai, Verdu, yet applies. To clarify, we limit the

---
[1]The I-MMSE identity in [1] with a derivative of the mutual information that is equal to half of the MMSE is attributed to real Gaussian noise.

---

exploitation of Palomar, Verdu's result to the derivative with respect to the SNR also, therefore,

$$\frac{dI(snr)}{dsnr} = Tr\left\{\mathbf{H}_k \mathbf{P}_k \mathbf{E}_k (\mathbf{H}_k \mathbf{P}_k)^{\dagger}\right\} = mmse(snr) \quad (5)$$

The above result in (5) just shows a scaling to (3). Therefore, the two results in [1] and [3] are almost similar reflections of the same result. Those results have treated point-to-point scalar and linear vector Gaussian channels, respectively. Thus, in both cases the existence of interference is neglected. In other words, the interference is not interpreted in its two fundamental parts, which are the mutual interference (constructive) part and the non-mutual interference (destructive) part. More specifically, in this paper, we make it clear how the existence of a set of vector Gaussian channels interfering with another set can in effect changes the information rates changes of each.

We present the main contributions in the next sections, and how this work provides a generalized identity to ones for point-to-point, scalar, vector parallel Gaussian channels, and vector MIMO Gaussian channels [1], [3]. Additionally, we show how our derived generalized Multiuser I-MMSE identity, extends the Multiuser I-MMSE in [19] not only to K-users case, but it provides insights on a Multiterminal I-MMSE or a Network I-MMSE version [20] that discusses changes in the information flow rates across multiple terminals and nodes in a noisy networks.

## III. GENERALIZED $K$-USER I-MMSE

The first contribution is given in the following theorem, which provides a generalization of the I-MMSE identity to the $K$-users case.

*Theorem 1:* The relation between the derivative of the multiuser mutual information with respect to the SNR and the non-linear MMSE for a multiuser Gaussian channel satisfies:

$$\frac{dI(snr)}{dsnr} = mmse(snr) + \psi(snr) \quad (6)$$

Where,

$$mmse(snr) = \sum_{k=1}^{K} Tr\left\{\mathbf{H}_k \mathbf{P}_k \mathbf{E}_k (\mathbf{H}_k \mathbf{P}_k)^{\dagger}\right\} \quad (7)$$

$$\psi(snr) = -\sum_{k=1}^{K}\sum_{j=1, j\neq k}^{K} Tr\{\mathbf{H}_k \mathbf{P}_k \mathbb{E}_{\mathbf{y}}[\mathbb{E}_{\mathbf{x}_k|\mathbf{y}}[\mathbf{x}_k|\mathbf{y}]\mathbb{E}_{\mathbf{x}_j|\mathbf{y}}[\mathbf{x}_j|\mathbf{y}]^{\dagger}] \times$$
$$(\mathbf{H}_j \mathbf{P}_j)^{\dagger}\} \quad (8)$$

*Proof:* The proof is provided in [19]. ∎

The per-user MMSE is given as follows:

$$\mathbf{E}_k = \mathbb{E}_{\mathbf{y}}[(\mathbf{x}_k - \widehat{\mathbf{x}}_k)(\mathbf{x}_k - \widehat{\mathbf{x}}_k)^{\dagger}] \quad (9)$$

The non-linear input estimates of each user input is given as follows:

$$\widehat{\mathbf{x}}_k = \mathbb{E}_{\mathbf{x}_k|\mathbf{y}}[\mathbf{x}_k|\mathbf{y}]$$
$$= \sum_{\mathbf{x}_1,..,\mathbf{x}_K} \frac{\mathbf{x}_k p_{\mathbf{y}|\mathbf{x}_1,..,\mathbf{x}_K}(\mathbf{y}|\mathbf{x}_1,..,\mathbf{x}_K) \prod_{k=1}^{K} p_{\mathbf{x}_k}(\mathbf{x}_k)}{p_{\mathbf{y}}(\mathbf{y})} \quad (10)$$

For such a system, the conditional probability distribution of the Gaussian noise is defined as:

$$p_{\mathbf{y}|\mathbf{x}_1,\mathbf{x}_2,..,\mathbf{x}_K}(\mathbf{y}|\mathbf{x}_1,\mathbf{x}_2,..,\mathbf{x}_K) = \frac{1}{\pi^{n_r}} e^{-\|\mathbf{y} - \sum_{k=1}^{K} \sqrt{snr}\mathbf{H}_k\mathbf{P}_k\mathbf{x}_k\|^2} \quad (11)$$

Additionally, the probability density function for the received vector $\mathbf{y}$ is defined as:

$$p_y(\mathbf{y}) = \sum_{\mathbf{x}_1,\mathbf{x}_2,..,\mathbf{x}_K} p_{\mathbf{y}|\mathbf{x}_1,..,\mathbf{x}_K}(\mathbf{y}|\mathbf{x}_1,..,\mathbf{x}_K) \prod_{k=1}^{K} p_{\mathbf{x}_k}(\mathbf{x}_k). \quad (12)$$

Note that the term $mmse(snr)$ is due to the $K$-users MMSEs, particularly,

$$mmse(snr) = \sum_{k=1}^{K} mmse_k(snr) \quad (13)$$

and $\psi(snr)$ are covariance terms that appear due to the covariance of the $K$-users interference. In particular, this term corresponds to the interference of $K-1$ users to one user, and how each user contribute with an interference term to the other $K-1$ users. Those terms are with respect to the channels, precoders, and non-linear estimates of the user inputs as will be precisely provided in the next section.

*A. The $K$-User I-MMSE generalizes the I-MMSE identity by Guo, Shamai, Verdu*

In this subsection, we show that the result here of a Multiuser I-MMSE generalizes the I-MMSE by Guo, Shamai, Verdu result in [1]. When the covariance terms vanish to zero, the derivative of the mutual information with respect to the SNR will be equal to the MMSE with respect to the SNR. This applies to the relation for point to point communications or in other words if no interference is encountered by any user. Therefore, the result of Theorem 1 is a generalization of such connection between the two canonical operational measures in information theory and estimation theory - the mutual information and the MMSE - and boils down to the result of Guo et. al, [1] under certain conditions which are: (i) when the cross correlation between the inputs estimates equals zero (ii) when interference can be neglected. Therefore, when the term $\psi(snr)$ equals zero. The derivative of the mutual information with respect to the SNR equals the total $mmse(snr)$:

$$\frac{dI(snr)}{dsnr} = mmse(snr) \quad (14)$$

which matches the result by Guo et. al in [1].

Such generalized fundamental relation between the change in the multiuser mutual information and the SNR is of particular relevance. Firstly, it allows understanding the behavior of per-user rates with respect to the interference due to the mutual interference and the interference of other $K-1$ users in terms of their power levels and channel strengths. In addition, the result allows us to be able to quantify the losses incurred due to the interference in terms of bits. Moreover, this new relation yet captures the mutual interference introduced if cooperation is considered among nodes. Thus, this generalization is generic for point-to-point, and linear vector Gaussian channels. Extended version can be found in [19].

*B. The $K$-User I-MMSE generalizes the I-MMSE identity by Palomar-Verdu*

In this subsection, we show that the result of a Multiuser I-MMSE generalizes the I-MMSE by Palomar, Verdu result in [3] as well. Again, under the setups discussed in the previous subsection, the result of this work boils down to the one in [3] when $\psi(snr) = 0$.

Even for the MAC Gaussian channel with scalar channels per user, the gap from the cut-set bound exists almost surely. This corresponds to non-mutual interference that stands as a limiting factor to achieving capacity. However, such gap can be accounted for via adding some mutual interference that is equal to $-\psi(snr)$, thus negative terms are canceled with additive positive terms. In turn, an interference channel is moved into a MIMO channel via cooperation.

In the linear vector MIMO Gaussian channel discussed by Palomar and Verdu, the gradient forms of the I-MMSE was devised, however, the $\psi(snr)$ term does not exist. This is due to the fact that it is accounted for, due to cooperation, the term $\psi(snr)$ is canceled, and we are left with the gradient of the mutual information connected to the MMSE matrix only. This perspective of mutual and non-mutual interference interpretation was not available in their work. Thus, in [7] and [8], the term $\psi(snr)$ was provided to interpret the interference within devised I-MMSE gradients of the MAC Gaussian channel.

This means that [3], did not provide an interpretation of the interference. Additionally, the connection between conditional and non-conditional components of the users rates and the distance between both rates driven by the interference was completely absent in previous works. This distance defines clearly the gap from the cut set upper bound with respect to the channel, precoding (power allocation) and the inputs estimates, see [19]. Therefore, the effect of the type of estimation and order of estimation, and so the scaling effects on the SNR was absent as well. In turn, the result of this paper a generalization of both results in [1] and [3].

The generalized I-MMSE is extended beyond to allow for understanding interference leakage when inputs are decoded successively. The next section addresses such components in the information rates.

## IV. $K$-USER CONDITIONAL AND NON-CONDITIONAL I-MMSE WITH SIC DECODING

We capitalize on the new fundamental relation to extend the derivative with respect to the SNR to $K$-user conditional and non-conditional mutual information in the SIC process. Capitalizing on the chain rule of the mutual information, the joint mutual information of $K$-users is given as follows:

$$I(\mathbf{x}_1, \mathbf{x}_2, .., \mathbf{x}_K; \mathbf{y}) = I(\mathbf{x}_1; \mathbf{y}) + I(\mathbf{x}_2; \mathbf{y}|\mathbf{x}_1) + I(\mathbf{x}_3; \mathbf{y}|\mathbf{x}_1, \mathbf{x}_2)$$
$$+ ... + I(\mathbf{x}_K; \mathbf{y}|\mathbf{x}_1, \mathbf{x}_2, .., \mathbf{x}_{K-1}) \quad (15)$$

Therefore, through this observation we can conclude the following theorem.

*Theorem 2:* The relation between the derivative of the $K$-user conditional and the non-conditional mutual information and their corresponding minimum mean squared error, for a step by step SIC decoding, satisfies, respectively:

$$\frac{dI(\mathbf{x}_1;\mathbf{y})}{dsnr} = mmse_1(\gamma_1^{-1}snr) \quad (16)$$

$$\frac{dI(\mathbf{x}_2;\mathbf{y}|\mathbf{x}_1)}{dsnr} = mmse_2(\gamma_2^{-1}snr) + \psi_{1,2}(\gamma_2^{-1}snr) + \psi_{2,1}(\gamma_2^{-1}snr) \quad (17)$$

$$...$$

$$\frac{dI(\mathbf{x}_k;\mathbf{y}|\mathbf{x}_1,\mathbf{x}_2,..,\mathbf{x}_{K-1})}{dsnr} = mmse_K(\gamma_K^{-1}snr) + \psi_{1,K}(\gamma_K^{-1}snr) + \psi_{K,1}(\gamma_K^{-1}snr) \quad (18)$$

Where the total MMSE is given by,

$$mmse(snr) = \sum_{k=1}^{K} mmse_k(\gamma_k^{-1}snr) \quad (19)$$

The covariances due to $K-1$ interferers are given by,

$$\psi_{k,1}(\gamma_k^{-1}snr) = -\sum_{\ell=k}^{1} Tr\{\mathbf{H}_k\mathbf{P}_k\mathbb{E}_{\mathbf{y}}[\mathbb{E}_{\mathbf{x}_k|\mathbf{y}}[\mathbf{x}_k|\mathbf{y}]\mathbb{E}_{\mathbf{x}_\ell|\mathbf{y}}[\mathbf{x}_\ell|\mathbf{y}]^\dagger](\mathbf{H}_\ell\mathbf{P}_\ell)^\dagger\} \quad (20)$$

and,

$$\psi_{1,k}(\gamma_k^{-1}snr) = -\sum_{\ell=1}^{k} Tr\{\mathbf{H}_k\mathbf{P}_k\mathbb{E}_{\mathbf{y}}[\mathbb{E}_{\mathbf{x}_\ell|\mathbf{y}}[\mathbf{x}_\ell|\mathbf{y}]\mathbb{E}_{\mathbf{x}_k|\mathbf{y}}[\mathbf{x}_k|\mathbf{y}]^\dagger](\mathbf{H}_k\mathbf{P}_k)^\dagger\} \quad (21)$$

*Proof:* Details of the proof follows from Theorem 1. Notice that, $\gamma_k, k = 1, ..., K$ are scaling factors. note also that $\gamma_1 > \gamma_2 > \gamma_3 > ... > \gamma_K$, where $\gamma_K = 1$. Then, taking the derivative of both sides of (15), and subtracting the derivative of $I(\mathbf{x}_1; \mathbf{y})$ which is equal to user 1 $mmse_1(\gamma_1^{-1}snr)$, $\gamma_1$ is a scaling factor, due to the fact that $\mathbf{x}_1$ is decoded first considering the other $K-1$ users' inputs as noise. Then, subtracting the derivative of $I(\mathbf{x}_2; \mathbf{y}|\mathbf{x}_1)$ which corresponds to user 2 $mmse_2(\gamma_2^{-1}snr)$, $\gamma_2$ is a scaling factor, due to the fact that $\mathbf{x}_2$ is decoded second considering the other $K-2$ users' (except user 1) inputs as noise and based on the knowledge of $\mathbf{x}_1$, we have the added covariance $\psi_{1,2}(snr) + \psi_{2,1}(snr)$, where $\psi_{1,2}$ appears as interpretation of the user 2 interference on user 1, and $\psi_{2,1}(snr)$ appears as interpretation of the first user interference on user 2. Repeating the same steps, until user $K$, we have the derivative of $I(\mathbf{x}_k; \mathbf{y}|\mathbf{x}_1, \mathbf{x}_2, .., \mathbf{x}_{K-1})$ equals to a non-scaled $mmse_K(snr)$ plus the covariance $\psi_{1,2,..,K-2,K}(snr) + \psi_{K,K-1,K-2,..,1}(snr)$ caused by the $K$-th user interference on the other $K-1$ users and the $K-1$ users interference on the $K$-th user respectively. And $\psi(snr) = \sum_{\forall\{k,\ell\}\in\{1,...,K\}} \psi_{k,\ell}(snr)$ which corresponds to $K!$ covariance terms. Therefore, Theorem 2 has been proved, and matches with Theorem 1 where the sum of the derivatives of the per users' mutual information equals the derivative of the joint mutual information. ■

Of particular relevance is the implication of the derived relations on understanding the achievable rates of interference channels. In particular, such relation allows for better understanding of the changes in the rates due the interferer which is either decoded first or considered as noise in a SIC form. Additionally, the theorem can be analytically interpreted in a different way if we consider that part of the users consider partial set of the users interference as noise. Moreover, if a joint decoding process is in place, the interpretation again changes, just by moving some covariance terms systematically and removing the scaling of the MMSE.

In particular, such generalization explains in explicit steps how a one step ahead in the knowledge process via conditioning provides less error, thus more information rates of a user decoded one step ahead. Besides, looking back one step into the successive decoding or interference cancellation process, the relation provides explicit closed forms of the losses encountered due to being blind about other aspects of the process. Therefore, the scaling with larger variance makes the user away from the ultimate knowledge of the process. In turn, such generalization goes beyond the known I-MMSE by Guo, Shamai, Verdu [1] into a general relation that provides clear connection to the I-MMSE with lookahead and lookback in [21], even if continuous Gaussian channels were exploited. However, we limit the exploitation here to the discrete case.

Therefore, we can understand this result with the SIC process from another side as an application that clarifies the difference in the one step lookback $-d$ and lookahead $d$ MMSE and their corresponding losses and gains in the information rates due to scaling of the SNR and the change in the cross correlation between input estimates. Thus, the information measure connected to looking back or looking forward is not equal, i.e., $I(snr; \mathbf{y}_0^{-d}|\mathbf{y}_\infty^0) \neq I(snr; \mathbf{y}_0^d|\mathbf{y}_{-\infty}^0)$, respectively. Where $\mathbf{y}_{t_i}^{t_f}$ corresponds to an observation window $[t_i - t_f]$ in the time of process $\mathbf{y}$. To understand further this result more deeply and analytically, it is worth to note that the distance between a scaled knowledge and a precise one characterizes the amount of information loss. Technically, writing the distance between the derivatives of the conditional and non-conditional mutual information is,

$$\frac{dI(\mathbf{x}_k;\mathbf{y}|\mathbf{x}_1,...,\mathbf{x}_{k-1})}{dsnr} - \frac{dI(\mathbf{x}_k;\mathbf{y})}{dsnr} = mmse_k(\gamma_k^{-1}snr) - mmse_k(\gamma_k^{-1}snr) + \psi_{1,k}(\gamma_k^{-1}snr) + \psi_{k,1}(\gamma_k^{-1}snr) \quad (22)$$

In turn, the information rate gain due to one step ahead, or the information rate loss due to going one step back in the SIC is characterized by integrating (22),

$$I(\mathbf{x}_k;\mathbf{y}|\mathbf{x}_1,...,\mathbf{x}_{k-1}) - I(\mathbf{x}_k;\mathbf{y}) = \int \psi_{1,k}(\gamma_k^{-1}snr)dsnr + \int \psi_{k,1}(\gamma_k^{-1}snr)dsnr \quad (23)$$

## V. SPECIAL CASE OF THE I-MMSE: GAUSSIAN DISTRIBUTED INPUTS

In this subsection, we specialize our result to one interesting case. Particularly, when the inputs are Gaussian distributed. Applying an SIC framework in the previous section, the derivatives for the conditional and non-conditional mutual information closed forms known for Gaussian inputs follows the one in Theorem 2. However, the terms $\psi_{k,1}$ and $\psi_{1,k}$ will be zero due to linear estimation of the inputs driven by the linearity of the MMSE that relies on estimating the input given the other is noise and so on and so forth. Interested reader can refer to [22] to find a detailed proof of having zero gap $\psi(snr) = 0$ when the Gaussian inputs are perfectly estimated successively with perfect removal along the SIC process. Particularly, the orthogonality between the input estimates let the gap from the cut-set bound to be closed almost surely. Therefore, the generalized relation takes a form for SIC that scales the mmse(snr) components of each user. Thus it reads as,

$$\frac{dI(\mathbf{x}_1;\mathbf{y})}{dsnr} = mmse_1(\gamma_1^{-1} snr) \quad (24)$$

$$\frac{dI(\mathbf{x}_2;\mathbf{y}|\mathbf{x}_1)}{dsnr} = mmse_2(\gamma_2^{-1} snr) \quad (25)$$

$$\ldots$$

$$\frac{dI(\mathbf{x}_k;\mathbf{y}|\mathbf{x}_1,\mathbf{x}_2,..,\mathbf{x}_{K-1})}{dsnr} = mmse_K(\gamma_K^{-1} snr) \quad (26)$$

Where the non-conditional mutual information is as follows[2],

$$I(\mathbf{x}_1;\mathbf{y}) = \log\ \det\left(\mathbf{I} + \mathbf{H}_1\mathbf{P}_1\mathbf{P}_1^\dagger\mathbf{H}_1^\dagger\mathbf{\Gamma}_1^{-1}\right) \quad (27)$$

and the conditional mutual information is as follows,

$$I(\mathbf{x}_2;\mathbf{y}|\mathbf{x}_1) = \log\ \det\left(\mathbf{I} + \mathbf{H}_2\mathbf{P}_2\mathbf{P}_2^\dagger\mathbf{H}_2^\dagger\mathbf{\Gamma}_2^{-1}\right) \quad (28)$$

and in general, the conditional mutual information is given by,

$$I(\mathbf{x}_k;\mathbf{y}|x_1,...,\mathbf{x}_{k-1}) = \log\ \det\left(\mathbf{I} + \mathbf{H}_k\mathbf{P}_k\mathbf{P}_k^\dagger\mathbf{H}_k^\dagger\mathbf{\Gamma}_k^{-1}\right) \quad (29)$$

The scaling matrix corresponds to the covariance of the noise plus the interference caused by part of inputs which are not estimated yet, i.e. the ones in the conditioning have their interference effect precluded when estimated and perfectly removed. Therefore, it changes along the SIC process, such that,

$$\mathbf{\Gamma}_k = \left(\mathbf{I} + \sum_{i=k+1, i\neq k}^{K} \mathbf{H}_i\mathbf{P}_i\mathbf{P}_i^\dagger\mathbf{H}_i^\dagger\right) \quad (30)$$

In turn the MMSE matrices are in their usual linear form for Gaussian inputs and each has an SNR scaled with such noise plus interference covariance,

$$mmse_k(\gamma_k^{-1} snr) =$$
$$Tr\left\{\mathbf{H}_k\mathbf{P}_k\left(\mathbf{\Gamma_k} + \mathbf{H}_k\mathbf{P}_k\mathbf{P}_k^\dagger\mathbf{H}_k^\dagger\right)^{-1}(\mathbf{H}_k\mathbf{P}_k)^\dagger\right\} \quad (31)$$

---
[2]It is worth to note that the mutual information and the MMSE are written for the system model in (4) when the inputs are all Gaussian distributed. This applies also to the relations in (24)-(25) with scalar like interpretation on the SNR scaling part.

The last estimated input has its MMSE not scaled since $\mathbf{\Gamma}_K = \mathbf{I}$, or for the scalar case, the last estimated input has its MMSE not scaled since $\gamma_K = 1$.

## VI. FUTURE RESEARCH DIRECTIONS

In this section we emphasize on the importance of the generalized Multiuser I-MMSE to several research directions in information theory and beyond, and to solving open problems or to generalizing state of the art schemes in communications systems.

First, this result will allow for extending the known optimal Mercury/Waterfilling power allocation by Lozano, Tulino, and Verdu in [12] into a more generic "Oil/Mercury/Waterfilling" optimal power allocation [23] that includes terms in the power allocation which can account for interference leakage when SIC is used, or with optimization of joint mutual information and joint estimation [8]. Additionally, such result allows for proposed network coding schemes that are capacity achieving [22].

Furthermore, other research problems can capitalize on this unveiled relation. For instance, the capacity of wireless networks using a network I-MMSE version will be exploited [20], the capacity of the interference channel can be investigated. Re-establishing new schemes and operational regimes for multihop network information theory [22]. The design of interference aware schemes and network coding schemes, e.g. [22], [16] and the design of power allocation for non-orthogonal signaling e.g. [22] are foreseen. Additionally, artificial intelligence, statistical signal processing and compressive sensing can capitalize on such tool.

## VII. CONCLUSIONS

We generalize the fundamental relation between the mutual information and the MMSE - I-MMSE identity - to the $K$-user Gaussian MIMO MAC channel, we present the effect of SIC on the characterization of such result. Such generalization provides an application to the I-MMSE with one step lookahead and lookback via quantifying the rates obtained or lost looking ahead or back into the successive decoding process.